\documentclass[12pt]{iopart}
\usepackage{epsfig}
\begin{document}
\title{Microscopic many-body theory of atomic Bose
       gases near a Feshbach resonance}
\author{R A Duine and H T C Stoof}
\address{Institute for Theoretical Physics,
         University of Utrecht, Leuvenlaan 4,\\
         3584 CE Utrecht, The Netherlands}
\ead{\mailto{duine@phys.uu.nl}, \mailto{stoof@phys.uu.nl}}
\begin{abstract}
A Feshbach resonance in the $s$-wave scattering length occurs if
the energy of the two atoms in the incoming open channel is close
to the energy of a bound state in a coupled closed channel.
Starting from the microscopic hamiltonian that describes this
situation, we derive the effective atom-molecule theory for a
Bose gas near a Feshbach resonance. In order to take into account
all two-body processes, we have to dress the bare couplings of the
atom-molecule model with ladder diagrams. This results in a
quantum field theory that exactly reproduces the scattering
amplitude of the atoms and the bound-state energy of the
molecules. Since these properties are incorporated at the quantum
level, the theory can be applied both above and below the critical
temperature of the gas. Moreover, making use of the true
interatomic potentials ensures that no divergences are
encountered at any stage of the calculation. We also present the
mean-field theory for the Bose-Einstein condensed phase of the
gas.
\end{abstract}
\submitto{\JOB}
\maketitle

\def\bx{{\bf x}}
\def\bk{{\bf k}}
\def\half{\frac{1}{2}}
\def\args{(\bx,t)}
\def\argsp{(\bx',t)}
\def\psiup{\hat \psi_{\uparrow}}
\def\psidup{\hat \psi_{\uparrow}^{\dagger}}
\def\psidown{\hat \psi_{\downarrow}}
\def\psiddown{\hat \psi_{\downarrow}^{\dagger}}
\def\psim{\hat \psi_{\rm m}}
\def\psimd{\hat \psi_{\rm m}^{\dagger}}

\section{Introduction}
\label{sec:introduction} The recent experiment by Donley {\it et
al.} \cite{JILA1} has made it clear that near a Feshbach resonance
the coherence between atoms and molecules can have a profound role
in the dynamics of an atomic Bose-Einstein condensate. As a
result, it has become an urgent problem to understand from first
principles how to properly incorporate the possible coherence
between atoms and molecules into the theory of an interacting
Bose gas. In the last four years important progress towards a
solution of this problem has been made by a number of groups
\cite{peter,eddy,murray1,servaas,juha,keith}. Nevertheless, it
appears that a fully satisfactory theory, which obeys all the
requirements that on physical grounds can {\it a priori} be
imposed upon the theory, still needs to be developed. It is with
this goal in mind that the present contribution has been written.

A first requirement for the theory is that it is based on an
adequate microscopic description of a Feshbach resonance. Feshbach
resonances are already known for a long time in nuclear physics
\cite{feshbach}, but have only more recently been predicted to
occur in ultracold atomic gases \cite{stwalley,eite}. The
defining feature of such resonances is that they can only occur
in a multi-channel scattering problem. More precisely, a Feshbach
resonance occurs when the kinetic energy of the particles in the
incoming open channel is equal to the energy of a bound state in
a closed channel that is (weakly) coupled to the incoming
channel. It is important for our purposes that the physics of a
Feshbach resonance is quite different from the physics of a
resonance in a single-channel scattering problem with a bound
state near the continuum threshold of the particles. This can
most easily be seen from the fact that in the latter case the
bound state in the incoming channel is only very weakly bound and
the extent of its wave function is, therefore, always much larger
than the range of the interaction between the particles. For a
Feshbach resonance the extent of the bound state wave function is
generally of the same size as the range of the interactions,
because it usually corresponds to a deeply bound state in the
potential of another and closed channel. As a result also the
quantum numbers of the bound state are different from the quantum
numbers of the incoming particles, which is clearly not the case
for the single-channel scattering problem.

A second requirement follows from the fact that the many-body
theory based on the microscopic description envisaged above is
free of ultra-violet divergences at any level of approximation,
and in particular, for both the normal and superfluid phases of
the atomic Bose gas. The reason for this is that microscopically
the interatomic interactions responsible for the Feshbach
resonance are determined by short-range potentials, which cut off
all the momentum integrals that arise when we diagramatically
include the effects of the interactions. As a result the desired
effective atom-molecule hamiltonian of the gas can contain only
terms with coefficients that are finite and cut-off independent.
Moreover, the coefficients are determined by only a small number
of experimental parameters such as the position and the width of
the Feshbach resonance, for instance.

The most crucial additional requirement on the effective
atom-molecule hamiltonian is that it must exactly reproduce the
two-body physics of the atomic gas, {\it i.e.}, by solving the
hamiltonian in the Hilbert space of two atoms we must recover the
correct scattering amplitude of the atoms and also the correct
binding energy and quantum numbers of the molecule. Note that in
this manner the molecular properties are exactly incorporated at
the quantum level, which is important for several reasons. First,
the theory can now be applied both in the normal and in the
superfluid phase of the gas. Second, we are not restricted to a
mean-field description of the gas and are now also able to
systematically study fluctuation effects, which we have recently
shown to be of importance under the experimental conditions of
interest \cite{rembert}. Finally, the theory can be immediately
generalized to atomic Fermi gases near a Feshbach resonance,
which are of great current interest in view of the prospect of
creating new neutral BCS superfluids \cite{stoof3,murray2,allan}.
Also in this case fluctuation effects are known to be essential
\cite{combescot,chris} and can only be accounted for after the
molecular properties are exactly incorporated in the quantum
theory.

As already implicitly mentioned, the effective atom-molecule
hamiltonian is in first instance the most convenient way to arrive
at a mean-field description of the gas. In the case of a
Bose-Einstein condensed atomic gas, an exact property of the
system is that it has a gapless excitation. Therefore, any
physically reasonable mean-field theory must have the same
property. From a fundamental point of view the gapless excitation
is due to the fact that a Bose-Einstein condensate spontaneously
breaks the global $U(1)$ symmetry associated with the conservation
of the total number of atoms. For the mean-field theory to fulfill
this requirement automatically it must be formulated such that it
does not contain an anomalous density or pairing field
\cite{stoof1,nick}. Including an anomalous density in general
leads to a double counting of the interaction effects and,
therefore, destroys the gaplessness of the mean-field theory.

In agreement with the above discussion, the paper is organized as
follows. In \sref{sec:baretheory} we start from a microscopic
hamiltonian for an atomic Bose gas with a Feshbach resonance and
derive from this a bare atom-molecule theory. To make contact
with the experimentally known parameters of the Feshbach
resonance, we then carry out in \sref{sec:ladders} a complete
ladder summation to arrive at the desired effective atom-molecule
hamiltonian that incorporates exactly all the relevant two-atom
physics. In \sref{sec:boundstatedos} we show in particular how
the correct properties of the molecule are recovered. After that
we discuss in \sref{sec:meanfield} how to arrive at the simplest
mean-field theory that is appropriate at the low temperatures of
interest experimentally, where the thermal component of the gas
can be neglected. We finally end in \sref{sec:conclusions} with
our conclusions.

\section{Bare atom-molecule theory}
\label{sec:baretheory} Without loss of generality we can consider
the simplest situation in which a Feshbach resonance arises, {\it
i.e.}, we consider a homogeneous gas of identical atoms with two
internal states denoted by $|\uparrow \rangle$ and $|\downarrow
\rangle$. The atoms in the two states interact via
the potentials $V_{\uparrow \uparrow} ({\bf x}-{\bf x}')$ and
$V_{\downarrow \downarrow} ({\bf x}-{\bf x}')$, respectively. The
state $| \downarrow \rangle$ has an energy $\Delta \mu B/2$ with
respect to the state $|\uparrow \rangle$ due to the Zeeman
interaction with the magnetic field $B$. The coupling between the
two states, which from the atomic physics point of view is due to
the hyperfine interaction, is denoted by $V_{\uparrow \downarrow}
(\bx-\bx')$. Putting everything together our microscopic
hamiltonian is thus given by
\begin{eqnarray}
\label{eq:hamiltonian}
   &&\hat H = \int d {\bf x}
        \psidup (\bx) \left[
     -\frac{\hbar^2 {\bf
      \nabla}^2}{2m}+ \half \int d \bx' \psidup (\bx')
      V_{\uparrow \uparrow} (\bx-\bx') \psiup (\bx')
    \right] \psiup (\bx) \nonumber \\
      &&+ \int d {\bf x}
        \psiddown (\bx) \left[
     -\frac{\hbar^2 {\bf
      \nabla}^2}{2m} +\frac{\Delta \mu B}{2}
                     + \half \int d \bx' \psiddown (\bx')
      V_{\downarrow \downarrow} (\bx-\bx') \psidown (\bx')
    \right] \psidown (\bx) \nonumber \\
      &&+\half \int d \bx \int d \bx' \left[ \psidup (\bx) \psidup (\bx')
         V_{\uparrow \downarrow} (\bx-\bx') \psidown (\bx') \psidown (\bx)
         + {\rm
     h.c.} \right]~,
\end{eqnarray}
where the potential $V_{\downarrow\downarrow} (\bx-\bx')$ is
assumed to contain the bound state responsible for the Feshbach
resonance. Using a Hubbard-Stratonovich transformation to
decouple this part of the hamiltonian \cite{kleinert,stoof2}, we
introduce the molecular field operator $\psim (\bx)$ that
annihilates a molecule at position $\bx$. In the approximation
that we are close to resonance, only a single bound state
contributes and this operator has the property that
\begin{equation}
  \langle  \psidown (\bx) \psidown (\bx') \rangle
  = \sqrt{2} \langle \hat \psi_{\rm m}((\bx+\bx')/2) \rangle
                 \chi_{\rm m} (\bx-\bx').
\end{equation}
The properly normalized and symmetrized bound state wave function
in the potential $V_{\downarrow\downarrow} (\bx-\bx')$, which we
choose to be real for simplicity, obeys the Schr\"odinger equation
\begin{equation}
  \left[ -\frac{\hbar^2 {\bf \nabla}^2}{m}
   + V_{\downarrow\downarrow} (\bx)
  \right] \chi_{\rm m} (\bx) = E_{\rm m} \chi_{\rm m}
  (\bx).
\end{equation}
After the Hubbard-Stratonovich transformation we obtain the bare
hamiltonian for the coupled atom-molecule system. It reads
\begin{eqnarray}
\label{eq:atommoleham}
   &&\hat H = \int d {\bf x}
        \psidup (\bx) \left[
     -\frac{\hbar^2 {\bf
      \nabla}^2}{2m}+ \half \int d \bx' \psidup (\bx')
      V_{\uparrow \uparrow} (\bx-\bx') \psiup (\bx')
    \right] \psiup (\bx) \nonumber \\
      &&+ \int d {\bf x}
        \psimd (\bx) \left[
     -\frac{\hbar^2 {\bf
      \nabla}^2}{4m} +\Delta \mu B + E_{\rm m}
    \right] \psim (\bx) \nonumber \\
      &&+\int d \bx \int d \bx'
        \left[ g_{\uparrow\downarrow}(\bx-\bx') \psimd ((\bx+\bx')/2)
              \psiup (\bx') \psiup (\bx) + {\rm
     h.c.} \right]~,
\end{eqnarray}
where $g_{\uparrow\downarrow}(\bx)=V_{\uparrow\downarrow} (\bx)
\chi_{\rm m} (\bx)/\sqrt{2}$ is the bare atom-molecule coupling.
The molecule-molecule and atom-molecule interactions also follow
from the above procedure but will be neglected in the following,
since under the experimental conditions of interest
\cite{JILA1,JILA2} the density of molecules is very small.

\section{Ladder summations}
\label{sec:ladders} For an the application of the hamiltonian in
\eref{eq:atommoleham} to realistic atomic gases we have to include
all two-body processes, because at the relevant low densities
three and more-body processes can in first instance be neglected.
These are most conveniently included by a renormalization of the
bare potential $V_{\uparrow \uparrow} (\bx-\bx')$ and the bare
coupling $g_{\uparrow\downarrow}(\bx)$. Moreover, the molecules
acquire a self energy.

\begin{figure}
\begin{center}
\epsfbox{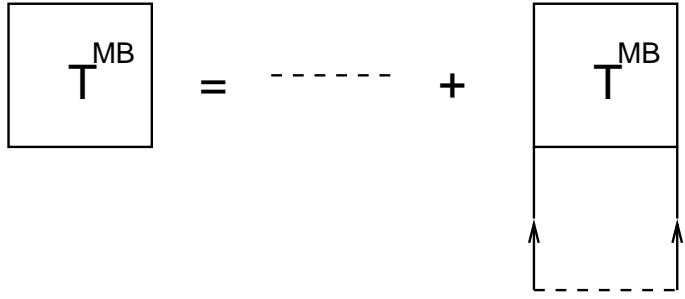}
\end{center}
\caption{\label{fig:tmatrix} Diagrammatic representation of the
many-body T matrix. The solid lines correspond to single-atom
propagators. The dashed lines corresponds to the interatomic
interaction $V_{\uparrow\uparrow}$.}
\end{figure}

The interaction potential of the atoms in principle renormalizes
to the many-body T(ransition) matrix. This renormalization is
determined by a Bethe-Salpeter equation which, within the for our
purposes sufficiently accurate Hartree-Fock approximation
\cite{stoof1}, reads
\begin{eqnarray}
\label{eq:mbtmatrix}
  T^{\rm MB} (\bk,\bk',{\bf K},z) &=& V_{\uparrow \uparrow} (\bk-\bk')
   \nonumber \\ && + \frac{1}{V} \sum_{\bk''}
    \frac{\left[ 1+ N(\epsilon_{{\bf K}/2+\bk''}-\mu)
                  + N(\epsilon_{{\bf K}/2-\bk''}-\mu)\right]}
         {z-\epsilon_{{\bf K}/2+\bk''}-\epsilon_{{\bf K}/2-\bk''}}
     \nonumber \\  && \times T^{\rm MB} (\bk'',\bk',{\bf K},z)~,
\end{eqnarray}
where $N(x)=[e^{\beta x}-1]^{-1}$ is the Bose distribution
function of the atoms, $\mu$ their chemical potential, and $1/k_{
\rm B}T$ the inverse thermal energy. This equation describes the
scattering of a pair of atoms from relative momentum $\bk'$ to
relative momentum $\bk$ at energy $z$. Due to the fact that the
scattering takes places in a medium the many-body T matrix also
depends on the center-of-mass momentum ${\bf K}$. The energy of a
single atom is equal to $\epsilon_{\bk}=\hbar^2 \bk^2/2m$. In the
Hartree-Fock approximation the energy in principle also contains
a mean-field correction which we neglect in first instance. We
come back to this point in \sref{sec:boundstatedos}, however. The
diagrammatic representation of \eref{eq:mbtmatrix} is given in
\fref{fig:tmatrix}.

For temperatures not too close to the critical temperature we are
allowed to neglect the many-body effects \cite{stoof1}, and
\eref{eq:mbtmatrix} reduces to the Lippmann-Schwinger equation for
the two-body T matrix. The effective interaction between the atoms
thus becomes $T^{\rm 2B}({\bf k},{\bf k}',z-\epsilon_{\bf K}/2)$. For
the realistic conditions of the atomic gases under consideration
here, {\it i.e.}, small external momenta and energies, the
two-body T matrix is independent of momentum and energy and equal
to $4 \pi a_{\uparrow\uparrow}\hbar^2/m$, with
$a_{\uparrow\uparrow}$ the $s$-wave scattering length of the
potential $V_{\uparrow\uparrow}(\bx-\bx')$. Therefore, we conclude
that the renormalization of this potential is given by
\begin{equation}
  V_{\uparrow \uparrow} (\bx-\bx') \to \frac{4 \pi
  a_{\uparrow\uparrow}\hbar^2}{m} \delta (\bx-\bx')~.
\end{equation}

\begin{figure}
\begin{center}
\epsfbox{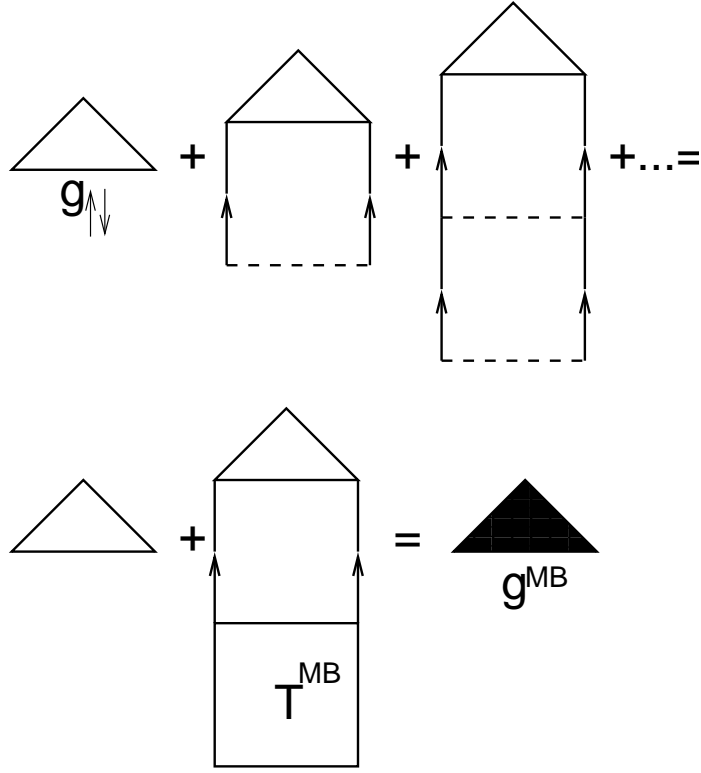}
\end{center}
\caption{\label{fig:gren} Diagrammatic representation of
renormalization of the bare atom-molecule coupling
$g_{\uparrow\downarrow}$.}
\end{figure}

The renormalization of the bare atom-molecule coupling is determined by the
equation
\begin{eqnarray}
\label{eq:rengb}
  g^{\rm MB} (\bk,{\bf K},z) &=& g_{\uparrow\downarrow}(\bk)
  + \frac{1}{V} \sum_{\bk'}
   T^{\rm MB} (\bk,\bk',{\bf K},z) \nonumber \\
   &&\times \frac{\left[ 1+ N(\epsilon_{{\bf K}/2+\bk'}-\mu)
                          + N(\epsilon_{{\bf K}/2-\bk'}-\mu)\right]}
                 {z-\epsilon_{{\bf K}/2+\bk'}-\epsilon_{{\bf K}/2-\bk'}}
    g_{\uparrow\downarrow}(\bk')~,
\end{eqnarray}
and is presented diagramatically in \fref{fig:gren}. Neglecting
many-body effects, the coupling constant becomes $g^{\rm
2B}(\bk,z-\epsilon_{\bf K}/2)$ with
\begin{eqnarray}
\label{eq:rengb2b}
  g^{\rm 2B} (\bk,z) = g_{\uparrow\downarrow}(\bk) + \frac{1}{V} \sum_{\bk'}
   T^{\rm 2B} (\bk,\bk',z) \frac{1}{z-2\epsilon_{\bf k'}}
    g_{\uparrow\downarrow}(\bk')~.
\end{eqnarray}
For the relevant small momenta and energies we are thus lead to the
substitution
\begin{equation}
  g_{\uparrow\downarrow}(\bx-\bx') \to g \delta (\bx-\bx')~,
\end{equation}
where $g$ can be related to experimentally known parameters as
follows. The resonance is characterized experimentally by a width
$\Delta B$ and a position $B_0$. More precisely, the $s$-wave
scattering length of the atoms as a function of magnetic field is
given by
\begin{equation}
\label{eq:ascatofb}
  a(B) = a_{\rm bg} \left( 1-\frac{\Delta B}{B-B_0} \right),
\end{equation}
where $a_{\rm bg}$ denotes the so-called background scattering
length. To make correspondence with the experiment we thus have
that $a_{\uparrow\uparrow}=a_{\rm bg}$. In order to reproduce the
experimentally observed width of the resonance we have that
$g=\hbar \sqrt{2 \pi a_{\rm bg} \Delta B  \Delta \mu /m}$, since
an elimination of the molecular field shows that $a_{bg} \Delta B
= m g^2/(2 \pi \hbar^2 \Delta \mu)$.

\begin{figure}
\begin{center}
\epsfbox{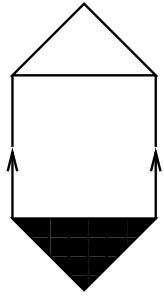}
\end{center}
\caption{\label{fig:selfenergy} Diagrammatic representation of the self
energy of the molecules.}
\end{figure}

The self energy of the molecules, shown diagramatically in
\fref{fig:selfenergy}, is given by
\begin{eqnarray}
\label{eq:selfenergy}
  \hbar \Sigma^{\rm MB} ({\bf K},z) &=& \frac{2}{V}
      \sum_{\bk} g_{\uparrow\downarrow}(\bk)
      \frac{\left[ 1+ N(\epsilon_{{\bf K}/2+\bk}-\mu)+N(\epsilon_{{\bf
    K}/2-\bk}-\mu)\right]}{z-\epsilon_{{\bf K}/2+\bk}-\epsilon_{{\bf K}/2-\bk}}
    \nonumber \\ && \times g^{\rm MB} (\bk,{\bf K},z)~.
\end{eqnarray}
In first instance we neglect again many-body effects which reduces
the self energy in \eref{eq:selfenergy} to $\hbar \Sigma^{\rm 2B}
(z-\epsilon_{\bf K}/2)$ with
\begin{equation}
\label{eq:selfenergytwobody}
  \hbar \Sigma^{\rm 2B} (z) =  \langle
  \chi_{\rm m} | \hat V_{\uparrow \downarrow} \hat G_{\uparrow\uparrow} (z)
                 \hat V_{\uparrow \downarrow}|
  \chi_{\rm m} \rangle~,
\end{equation}
where the propagator $\hat G_{\uparrow\uparrow}(z)$ is given by
\begin{equation}
  \hat G_{\uparrow\uparrow}(z) = \frac{1}{z-\hat H_{\uparrow\uparrow}}~,
\end{equation}
with the hamiltonian
\begin{equation}
  \hat H_{\uparrow\uparrow} = \frac{\hat {\bf p}^2}{m}
                              + \hat V_{\uparrow \uparrow}
  \equiv \hat H_0 + \hat V_{\uparrow \uparrow}~.
\end{equation}
We insert in \eref{eq:selfenergytwobody} a complete set of bound
states $|\phi_{\kappa}\rangle$ with energies $E_\kappa$ and
scattering states $|\phi^{(+)}_{\bk} \rangle$. The latter obey the
Lippmann-Schwinger equation
\begin{equation}
\label{eq:lipschwstates}
  | \phi^{(+)}_{\bk} \rangle = | \bk \rangle + \frac{1}{2
  \epsilon_{\bk}^+ -\hat H_0} \hat V_{\uparrow \uparrow}
  | \phi^{(+)}_{\bk} \rangle~,
\end{equation}
where $\epsilon^+_{\bk}=\epsilon_{\bk}+i0$ denotes the usual
limiting procedure. This reduces the self energy $\hbar
\Sigma^{\rm 2B} (z)$ to
\begin{eqnarray}
\label{eq:selfenergyint}
  \hbar \Sigma^{\rm 2B} (z) &=&
    \sum_{\kappa} |\langle \chi_{\rm m}|
           \hat V_{\uparrow \downarrow}|\phi_{\kappa}\rangle |^2
    \frac{1}{z-E_{\kappa}} \\ \nonumber
&&  + \int \frac{d \bk}{(2 \pi)^3} |\langle \chi_{\rm m}|
           \hat V_{\uparrow \downarrow}|\phi^{(+)}_{\bk}\rangle |^2
    \frac{1}{z-2 \epsilon_\bk},
\end{eqnarray}
where we replaced the sum over the momenta {\bf k} by an integral.
Using \eref{eq:rengb2b} and the Lippmann-Schwinger equation we
have that
\begin{equation}
   g^{\rm 2B} (\bk, 2 \epsilon_{\bk}^+)=
  \frac{1}{\sqrt{2}} \langle \chi_{\rm
  m} |\hat V_{\uparrow \downarrow}|\phi^{(+)}_{\bk}\rangle~.
\end{equation}
As a result we have for the retarded self energy $\hbar
\Sigma^{(+)} (\hbar \omega)$, {\it i.e.}, the self energy
$\hbar\Sigma^{2B}(z)$ evaluated at the physically relevant energy
$z=\hbar \omega^+$ that
\begin{equation}
\label{eq:resultselfenergy}
  \hbar \Sigma^{(+)} (\hbar \omega) \simeq - g^2 \frac{m^{3/2}}{2 \pi \hbar^3} i
  \sqrt{\hbar \omega} - \left( \Delta \mu B_0 + E_{\rm m} \right),
\end{equation}
where we have denoted the energy-independent shift, that results from the
sum over bound states and the principal-value part of the integral in
\eref{eq:selfenergyint}, in such a manner that the position of the
resonance in the magnetic field is precisely at the experimentally
observed magnetic field value $B_0$.

\section{Molecular binding energy and density of states}
\label{sec:boundstatedos} Putting the results of the previous
sections together, we find that the Bose gas near a Feshbach
resonance is described by a coupled set of equations of motion for
the atomic and molecular Heisenberg operators
\begin{eqnarray}
  \hat \psi_{\rm a} (\bx,t)=\exp\left[\frac{i}{\hbar} \hat
  H t\right] \psiup (\bx) \exp\left[-\frac{i}{\hbar} \hat H t\right]~,
                                                             \nonumber \\
   \hat \psi_{\rm m} (\bx,t)=\exp\left[\frac{i}{\hbar} \hat
  H t\right] \psim (\bx) \exp\left[-\frac{i}{\hbar} \hat H t\right]~,
\end{eqnarray}
given by
\begin{eqnarray}
\label{eq:heom}
  i \hbar \frac{\partial \hat \psi_{\rm a} \args}{\partial t}
    =\left[  -\frac{\hbar^2 {\bf \nabla}^2}{2m}
            + T^{\rm 2B}_{\rm bg} \hat \psi_{\rm a}^{\dagger}
        \args \hat \psi_{\rm a} \args
    \right] \hat \psi_{\rm a} \args \nonumber \\ \qquad
    + 2 g \hat \psi_{\rm a}^{\dagger}
          \args \psim \args~, \nonumber \\
   i \hbar \frac{\partial \hat \psi_{\rm m} \args}{\partial t}
    =\left[  -\frac{\hbar^2 {\bf \nabla}^2}{4m}
            +\delta (B(t)) \right. \nonumber \\
    \qquad \left. - g^2 \frac{m^{3/2}}{2 \pi \hbar^3} i
  \sqrt{i \hbar \frac{\partial}{\partial t}
    +\frac{\hbar^2 {\bf \nabla}^2}{4m}}~
    \right] \hat \psi_{\rm m} \args + g \hat \psi_{\rm a}^2
    \args~,
\end{eqnarray}
where the detuning is defined by $\delta (B)=\Delta \mu (B-B_0)$
and $T^{\rm 2B}_{\rm bg} = 4 \pi a_{\rm bg} \hbar^2/m$. This is
the most important result of our work. Note that the
time-derivative and gradient terms appear exactly such that both
equations of motion are manifestly galilean invariant. Note also
that an external trapping potential can just be added to the
right-hand sides of these equations if required. It is interesting
to mention that the above equations can immediately be generalized
to a Fermi gas near a Feshbach resonance. Moreover, a simple
Hartree-Fock approximation to the resulting theory reproduces
exactly the interesting crossover physics recently discussed by
Ohashi and Griffin on the basis of the Nozi\`eres and Schmitt-Rink
formalism for the normal phase of the gas \cite{allan}. Having
made this observation, it is now clear how the same crossover
phenomena can be studied in the superfluid phase of the gas. Work
in this direction is in progress and will be reported elsewhere.

From \eref{eq:heom} we determine the retarded Green's function of the
molecules $G_{\rm m}^{(+)} (\bx,t;\bx',t')$. For fixed detuning, the poles of
its Fourier transform determine the bound-state energy. This Fourier
transform is given by
\begin{equation}
\label{eq:gmkw}
  G_{\rm m}^{(+)} (\bk,\omega)=
    \frac{\hbar}{\hbar \omega^+  - \epsilon_{\bk}/2 -\delta (B)
     + (g^2 m^{3/2}/2 \pi \hbar^3) i
     \sqrt{\hbar \omega - \epsilon_{\bk}/2}}~,
\end{equation}
with $\epsilon_{\bk}/2 = \hbar^2 \bk^2/4m$ the kinetic energy of a
molecule. For positive detuning the propagator has a pole with nonzero
imaginary part, which shows that the molecule has a finite lifetime in
this case. In first approximation the energy of the molecule is
$\epsilon_{\rm m}(B)+\epsilon_{\bk}/2$ with $\epsilon_{\rm m}(B) =
\delta(B)$ and its rate of decay equals $\Gamma_{\rm m}(B) = (g^2
m^{3/2}/4 \pi \hbar^4) \sqrt{\delta(B)}$.  For negative detuning the
molecular propagator has a real pole at $\hbar \omega = \epsilon_{\rm
m}+\epsilon_{\bk}/2$, where the bound state energy of the molecule is
given by
\begin{equation}
\label{eq:bse}
  \epsilon_{\rm m} (B)= \delta (B) + \frac{g^4 m^3}{8 \pi^2 \hbar^6}
   \left[\sqrt{1-\frac{16 \pi^2 \hbar^6}{g^4 m^3} \delta (B)} -1\right].
\end{equation}
Close to the resonance the bound-state energy is, using
\eref{eq:ascatofb}, found to be equal to
\begin{equation}
\label{eq:ebres}
  \epsilon_{\rm m}(B)  = - \frac{\hbar^2}{m [a(B)]^2}~.
\end{equation}
It is important to realize that this last equation, which is well-known
to be true for a weakly bound state in a single-channel scattering
problem \cite{sakurai}, has now thus been proven to be also valid for the case of
a multi-channel Feshbach resonance.

\begin{figure}
\begin{center}
\epsfbox{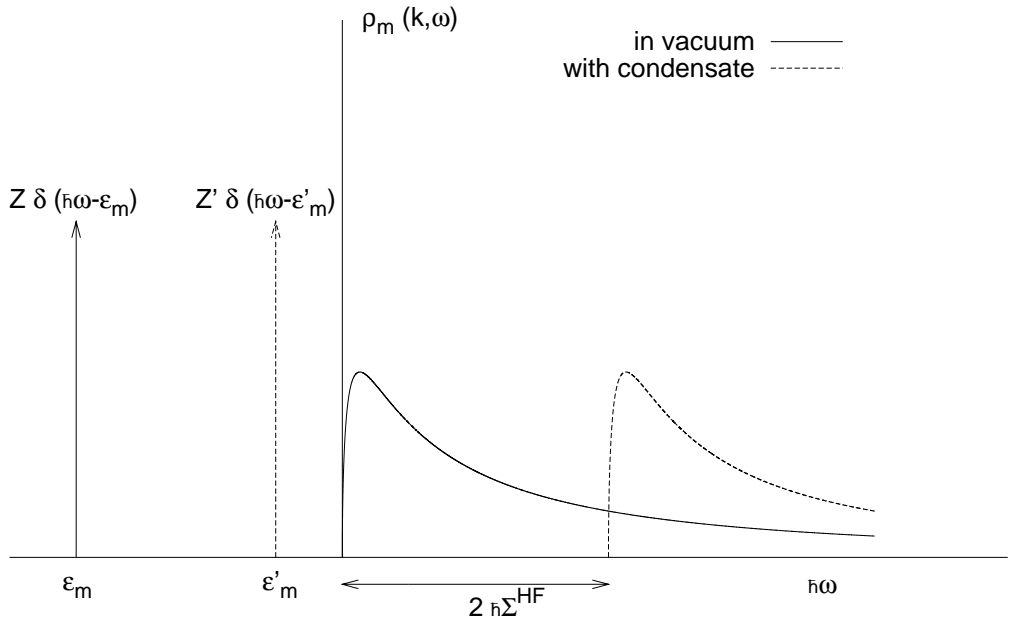}
\end{center}
\caption{\label{fig:dos} Molecular density of states.}
\end{figure}

The physics of \eref{eq:gmkw} is best understood by considering
the molecular density of states, related to the retarded propagator by
\begin{equation}
\label{eq:defdos}
  \rho_{\rm m} (\bk,\omega) = -\frac{1}{\pi}
               {\rm Im} \left[ G_{\rm m}^{(+)} (\bk, \omega) \right]~.
\end{equation}
For negative detuning, it has two contributions as shown in
\fref{fig:dos}. The first comes from the bound state, {\it i.e.}, the
(dressed) molecule, the second comes from the two-atom continuum. It is
the latter part of the density of states that incorporates into our theory
the rogue dissociation process put forward recently by Mackie {\it et al.}
\cite{juha}. More explicitly, the density of states is found to be equal
to
\begin{eqnarray}
\label{eq:dos}
  \rho_{\rm m} (\bk, \omega) =
   \frac{1}{1 + g^2 m^{3/2}/(4 \pi \hbar^3 \sqrt{|\epsilon_{\rm m}|})}
     \delta (\hbar \omega -\epsilon_{\bk}/2-\epsilon_{\rm m}) \nonumber \\
    ~+ \frac{1}{\pi} \theta (\hbar \omega - \epsilon_{\bk}/2)
      \frac{(g^2 m^{3/2}/2 \pi \hbar^3) \sqrt{\hbar \omega -
      \epsilon_{\bk}/2}}
      {\left[ \hbar \omega - \epsilon_{\bk}/2 - \delta (B) \right]^2
      +(g^4 m^3/4 \pi^2 \hbar^6)
      (\hbar \omega - \epsilon_{\bk}/2 )}~,
\end{eqnarray}
and can be shown to obey the sum rule
\begin{equation}
\label{eq:sumrule}
  \int d(\hbar\omega)~ \rho_{\rm m} (\bk,\omega) = 1~.
\end{equation}
We thus conclude that the wave function renormalization factor of the
molecules is given by $Z=1/[1 + g^2 m^{3/2}/(4 \pi \hbar^3
\sqrt{|\epsilon_{\rm m}|})]$, which close to resonance is much smaller than
one. Physically, this implies that in this case the wave function of the
molecule is strongly affected by the interaction with the continuum in the
incoming channel and contains only with an amplitude $\sqrt{Z}$ the wave
function $\chi_{\rm m} (\bx)$ of the bound state in the closed channel.

The molecular density of states changes when an atomic
Bose-Einstein condensate is present, due to the mean-field
interactions with the condensate. In particular, in this case we
expect that the two-atom continuum part of the density of states
is less important because of the mean-field barrier that the
two colliding condensate atoms have to overcome. Mathematically
this comes about since, to include the Hartree-Fock mean-field
shift of the energy of the atoms, we have to replace in
\eref{eq:gmkw} $\sqrt{\hbar\omega-\epsilon_{\bf k}/2}$ by
$\sqrt{\hbar\omega-2\hbar \Sigma^{\rm HF}-\epsilon_{\bf k}/2}$,
where $\hbar \Sigma^{\rm HF}$ denotes the Hartree-Fock
self energy of the noncondensed atoms due to their interaction
with the condensate.

\begin{figure}
\begin{center}
\epsfbox{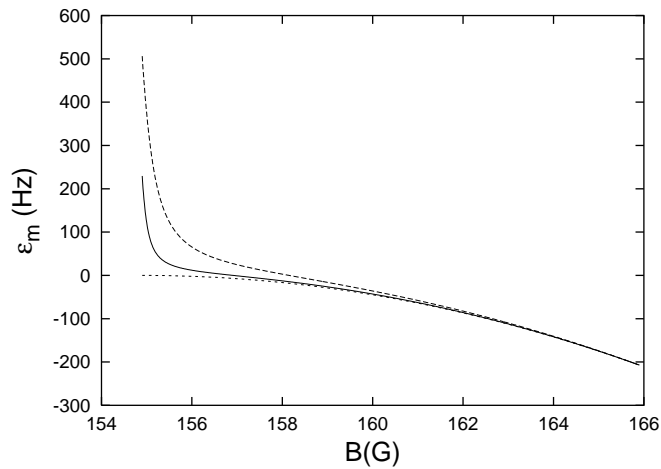}
\end{center}
\caption{\label{fig:boundstate} Mean-field shift of the
bound-state energy as a function of the magnetic field. The solid
line corresponds to an atomic condensate density of $n_0=1.1
\times  10^{13}$ cm$^{-3}$ and the dashed line corresponds to
$n_0=5.4 \times 10^{13}$ cm$^{-3}$. The dotted line corresponds
to the molecular binding energy in vacuum. The calculations are
performed for the resonance at $B_0 \simeq 154.9$ G in the
$|f=2;m_f=-2\rangle$ state of $^{85}$Rb \protect{\cite{JILA1,JILA2}}.}
\end{figure}

The mean-field shift of the thermal atoms leads to a change of the bound
state energy as well. In equilibrium we estimate the magnitude of this
shift by approximately calculating the self energy $\hbar \Sigma^{\rm HF}$
from
\begin{equation}
\label{eq:sigmahf}
   \hbar \Sigma^{\rm HF} \simeq
 2 n_0 \left[ T^{\rm 2B}_{bg} + \frac{2 g^2}{2 \hbar \Sigma^{\rm HF}-\delta (B)}
 \right]~,
\end{equation}
for a given condensate density $n_0$. In this manner we have calculated
the mean-field shift of the bound-state energy as a function of the
magnetic field for the Feshbach resonance at \mbox{$B_0 \simeq 154.9$ G}
in the $|f=2;m_f=-2\rangle$ state of $^{85}$Rb. In \fref{fig:boundstate}
the results of this calculation are presented for two experimentally
relevant condensate densities, namely $n_0=1.1 \times 10^{13}$ cm$^{-3}$
and $n_0=5.4 \times 10^{13}$ cm$^{-3}$ \cite{JILA1,JILA2}. As expected,
the shift of the bound-state energy is largest for the highest condensate
density and decreases away from the resonance. Although \eref{eq:sigmahf}
is only a first approximation to calculate the Hartree-Fock mean-field
energy, \fref{fig:boundstate} shows that the shift in the bound-state
energy is significant. A more thorough calculation of the mean-field
effect on the bound-state energy and also on the position of the resonance is
postponed to future work. With respect to this remark, it should be noted
that, since the retardation time of the interaction is only of the order
of $\hbar/\epsilon_{\rm m} (B)$, the mean-field effects are not instantaneous. Therefore,
the relevance of the mean-field shifts to nonequilibrium situations, such
as in the recent experiments with time-dependent detuning
\cite{JILA1,JILA2}, is also subject to further research.

\section{Mean-field theory}
\label{sec:meanfield} The mean-field theory for the coupled atomic and
molecular condensates is found by taking the expectation value of
\eref{eq:heom}. We assume for simplicity that the temperatures are so low
that the thermal cloud of the Bose gas can be neglected. It is,
however, straightforward to include the mean-field effects of the
thermal cloud in the same manner as in the by now standard Popov
theory for weakly-interacting Bose gases. Furthermore, we
neglect for simplicity the effect of the Hartree-Fock energy shift
on the two-atom continuum and the molecular binding energy. The
resulting mean-field equations are then given by
\begin{eqnarray}
\label{eq:mfe}
   i \hbar \frac{\partial \psi_{\rm a} (t)}{\partial t}
    = T^{\rm 2B}_{\rm bg} |\psi_{\rm a} (t)|^2
      \psi_{\rm a} (t) + 2 g  \psi_{\rm a}^* (t)
    \psi_{\rm m}(t)~, \nonumber \\
   i \hbar \frac{\partial\psi_{\rm m} (t)}{\partial t}
    =\left[ \delta (B(t)) - g^2 \frac{m^{3/2}}{2 \pi \hbar^3} i
  \sqrt{i \hbar \frac{\partial}{\partial t}}~
    \right]  \psi_{\rm m} (t) + g \psi_{\rm a}^2 (t)~.
\end{eqnarray}
Note that in agreement with our remarks in the introduction the
above equations contains no anomalous density or pairing field.
One way to understand the reason for this is that the effects of
the anomalous density are already included by using the
renormalized couplings and including the molecular self energy.
Including these effects again would lead to double counting
problems and, therefore, to a theory that is not gapless. Another
way to understand it is that equation \eref{eq:mfe} is explicitly
$U(1)$ invariant and the gaplessness of the theory is thus
automatically guaranteed.

A crucial ingredient in our formulation is the (nonlocal) term
proportional to
$\sqrt{i\hbar \partial/\partial t}$. It is this term that incorporates the
correct binding energy of the molecules for negative detuning, and their
lifetime if the detuning is positive. In principle this term must
be
treated as follows. Using the Green's function in \eref{eq:gmkw} we find
that the wave function of the molecular condensate is, for
time-independent detuning, given by
\begin{equation}
  \psi_{\rm m} (t) = \frac{g}{\hbar} \int_0^t d t' G^{(+)} (t-t') \psi_a^2
  (t')+\psi_{\rm m} (0)~,
\end{equation}
where the Fourier transform of $G^{(+)} ({\bf 0},\omega)$ is given by
\begin{eqnarray}
  G^{(+)} (t-t') =
   -i \theta (t-t') Z
     \exp \left[ -\frac{i}{\hbar} \epsilon_{\rm m} (t-t') \right] \nonumber \\
     \qquad  - \frac{i \theta (t-t') g^2 m^{3/2}}{\pi \hbar^3}
      \int_0^{\infty} \frac{d \omega}{2 \pi}
      \frac{\sqrt{\hbar \omega} e^{- i \omega (t-t')}}
      {\left[ \hbar \omega- \delta (B) \right]^2
      +(g^4 m^3/4 \pi^2 \hbar^6) \hbar \omega
      }~.
\end{eqnarray}
This result can then be substituted in the equation for the atomic
condensate wave function, which can now be easily solved
numerically as we will show in future work.

\section{Conclusions and outlook}
\label{sec:conclusions}
We have derived from first principles an effective quantum field
theory for Feshbach resonant interactions
in atomic Bose gases. In future work we intend to apply this
quantum field theory to
study various equilibrium and nonequilibrium properties of ultracold
atomic gases near a Feshbach resonance. This will include a
further study
of the mean-field shifts of the bound-state energy and the
position of the
resonance, as well as the study of the BEC/BCS crossover in the
superfluid phase of a two-component Fermi gas \cite{allan}.
Moreover, we are now in a position to also study the normal phase
for both for bosonic and fermionic gases.
In order to apply the theory also to the recent pulse experiments
with Bose-Einstein condensates of $^{85}$Rb \cite{JILA1,JILA2}, we
need to include a detuning that
varies rapidly with time. This leads to some technical
complications with the proper treatment of our nonlocal term that
remain to be resolved. With respect to the latter experiments we
also want to further study the importance of the quantum
evaporation process, which previous work has shown not to be
negligible \cite{rembert}.

\ack We would like to acknowledge the hospitality of the European
Centre for Theoretical Studies in Nuclear Physics and Related
Areas (ECT*) during the Summer Program on Bose-Einstein
condensation.

\end{document}